# Enhanced Visible Light Photocatalytic Activity for TiO$_2$ Nanotube Array Films by Codoping with Tungsten and Nitrogen

**Min Zhang, Juan Wu, DanDan Lu, and Jianjun Yang**

*Key Laboratory of Ministry of Education for Special Functional Materials, Henan University, Kaifeng 475004, China*

Correspondence should be addressed to Min Zhang; zm1012@henu.edu.cn and Jianjun Yang; yangjianjun@henu.edu.cn

A series of W, N codoped TiO$_2$ nanotube arrays with different dopant contents were fabricated by anodizing in association with hydrothermal treatment. The samples were characterized by scanning electron microscopy, X-ray diffraction, X-ray photoelectron spectroscopy, and ultraviolet-visible light diffuse reflection spectroscopy. Moreover, the photocatalytic activity of W and N codoped TiO$_2$ nanotube arrays was evaluated by degradation of methylene blue under visible light irradiation. It was found that N in codoped TNAs exists in the forms of Ti-N-O, while W exists as W$^{6+}$ by substituting Ti in the lattice of TiO$_2$. In the meantime, W and N codoping successfully extends the absorption of TNAs into the whole visible light region and results in remarkably enhanced photocatalytic activity under visible light irradiation. The mechanism of the enhanced photocatalytic activity could be attributed to (i) increasing number of hydroxyl groups on the surface of TNAs after the hydrothermal treatment, (ii) a strong W-N synergistic interaction leads to produce new states, narrow the band gap which decrease the recombination effectively, and then greatly increase the visible light absorption and photocatalytic activity; (iii) W ions with changing valences in all codoped samples which are considered to act as trapping sites, effectively decrease the recombination rate of electrons and holes, and improve the photocatalytic activity.

## 1. Introduction

Nowadays, photocatalysis has attracted lively interest due to its potential applications in environmental remediation and clean energy production. Among various photocatalysts, TiO$_2$ is most frequently employed owing to the advantages of earth abundance, low toxicity, and thermal and chemical stability. However, because of the wide band gap of titanium dioxide, only a small UV fraction of solar light (3–5%) can be utilized [1–3]. Many investigations have been conducted to extend optical absorption of TiO$_2$-based materials to the visible light region and to improve visible light photocatalytic activity by nonmetal doping using N [4], C [5, 6], S [7] or multielemental-doped materials [8–11]. In particular, TiO$_2$ photocatalysts codoped with N and another metal had attracted many attentions in recent years [12, 13]. For example, Yang et al. prepared C- and V-doped TiO$_2$ photocatalysts by a sol-gel method with high photocatalytic activity for the degradation of acetaldehyde both under visible light irradiation (>420 nm) and in the dark [14]. Liu et al. prepared a series of Ti$_{1-x}$Mo$_x$O$_{2-y}$N$_y$ samples using sol-gel method and found that Mo + N codoping can increase the up-limit of dopant concentration and create more impurity bands in the band gap of TiO$_2$ by first-principles band structure calculations [15]. It reveals that Mo + N codoped TiO$_2$ material is a promising photocatalyst with high photocatalytic activity under visible light.

Recently, many reports have focused on the W, N codoped TiO$_2$ nanoparticles to increase photocatalytic activity under visible light irradiation. Shen et al. fabricated the W, N codoped TiO$_2$ nanopowders using sol-gel and mechanical alloying methods. The samples exhibited strong absorbance in the visible range and enhanced photocatalytic activities under visible light irradiation from the results of photodegradation experiments and chemical oxygen demand analysis [16]. Li et al. prepared W, N codoped TiO$_2$ nanophotocatalysts with twist-like helix structure by a facile and template-free one-pot method, which showed higher visible light response compared with pure TiO$_2$ and P25 [17]. Kubacka et al. synthesized W- and N-doped TiO$_2$ anatase-based materials with both unprecedented high activity and selectivity in the gas-phase partial oxidation of aromatic hydrocarbons using sunlight as excitation energy and molecular oxygen as oxidant [18–20]. The doped tungsten and nitrogen species plays a key role in narrowing the band gap and then in expanding the photoactivity to visible light region and also in decreasing the recombination rate of excited electron and holes. However, most of the reported W- and N-doped TiO$_2$ photocatalysts were typically present in powders which usually display low surface area due to the closed packing and limit their industrial applications as photocatalysts.

Highly ordered TiO$_2$ nanotube arrays (TNAs) fabricated via electrochemical anodization of high purity Ti foils have attracted extensive interests in recent years. Owing to the presence of unique structural features such as large surface area, high porosity, and ordered pore channels facilitating reactant diffusion and adsorption, TiO$_2$ nanotube arrays are more attractive for use in the applications of photocatalysis [21], gas sensing [22], solar cells [23], photoelectrochemistry [24], and so on. However, full technological applications of TiO$_2$ nanotube arrays are hampered by the wide band gap of TiO$_2$ and the activation only under ultraviolet light irradiation. Although nonmetal doping of TiO$_2$ nanotube arrays has been confirmed as an effective method to enhance the photocatalytic activity under visible light irradiation [22, 23]; to our knowledge, there is little work about the fabrication and visible light photocatalytic activity of the W and N codoped TiO$_2$ nanotube arrays. In this work, we successfully prepared W, N codoped TiO$_2$ nanotube arrays by anodization and hydrothermal synthesis. The photocatalytic activity of as-prepared samples was studied based on methylene blue dye degradation under visible light irradiation. The results demonstrated that the W and N codoped TiO$_2$ nanotube arrays exhibited a higher photocatalytic activity than the single N-doped sample under visible light irradiation. The mechanism of visible light photocatalytic activity enhancement was also discussed.

## 2. Experimental Sections

*2.1. Synthesis of W and N Codoped TiO$_2$ Nanotube Arrays.* Self-organized and well-aligned TNAs were fabricated by two-step electrochemical anodization process [25]. Briefly, commercial titanium sheets (20 mm × 40 mm, purity > 99.6%) with a thickness of 0.25 mm were sequentially sonicated in acetone, isopropanol, and methanol for 10 min, followed by etching in the mixture of HF/HNO$_3$/H$_2$O for 20 s, rinsing with deionized water, and drying under a N$_2$ stream. Resultant rectangular Ti sheet was used as an anode to couple with Pt meshwork as a cathode in the anodic oxidation test setup. A direct current power supply and a mixed electrolyte solution of ethylene glycol containing 0.25 wt% NH$_4$F and 2 vol% deionized water were used in the electrochemical processes. The first oxidation step was conducted at 60 V for 1 h, with which the oxidized surface films were removed by sonication in distilled water before the oxidized samples were dried under high purity N$_2$ stream at room temperature. As-dried Ti substrates were subsequently oxidized in the original electrolyte at 60 V for 2 h, followed by sonication in ethanol for 7 min to remove the surface cover and drying under high purity N$_2$ stream to afford as-prepared TNAs. Magnetic stirring was adopted throughout the oxidation processes, while a circulation pump was performed at a low temperature of about 20°C. After electrochemical anodization, as-prepared TNAs were calcinated at 500°C for 3 h in a furnace. Interstitial nitrogen species were formed in the TNAs due to the electrolyte containing NH$_4$F [26]. We denoted these single N-doped TiO$_2$ nanotube arrays samples fabricated by two-step oxidization as N-TiO$_2$.

W and N codoped TNAs with different dopant concentration (0.01%, 0.05%, 0.5%, 1.00%, and 3.00%, referring to different concentration of W and N source) were then prepared by hydrothermally treating the anodized N-TiO$_2$ in a teflon-lined autoclave (120 mL, Parr Instrument) containing approximately 60 mL of ammonium tungstate ((NH$_4$)$_{10}$H$_2$(W$_2$O$_7$)$_6$) as the source of both W and N. The hydrothermal synthesis was conducted at 180°C for 5 h in a box oven, and then the autoclave was naturally cooled down to room temperature affording target products denoted as W-N-TiO$_2$-water, W-N-TiO$_2$-0.01, W-N-TiO$_2$-0.05, W-N-TiO$_2$-0.5, W-N-TiO$_2$-1, and W-N-TiO$_2$-3 (numeral suffixes water, 0.01, 0.05, 0.5, 1 and 3 refer to pure water and ammonium tungstate solution concentration of 0.01%, 0.05%, 0.50%, 1.00%, and 3.00%). All samples were rinsed with distilled water and dried under high purity N$_2$ stream at room temperature.

*2.2. Characterization.* Surface morphologies of TNAs were observed using a field emission scanning electron microscopy (FESEM, Nova NanoSEM 230). A Philips X'Pert Pro PW3040/60 X-ray diffractometer (XRD, Philips Corporation, Holland) was used to determine the crystal structure of as-prepared samples. Chemical state analysis was conducted with an Axis Ultra X-ray photoelectron spectroscopy (XPS, Kratos, UK; a monochromatic Al source operating at 36 W with a pass energy of 40 eV and a step of 0.1 eV was used). All XPS spectra were corrected using the C 1$s$ line at 284.8 eV. A Cary-5000 scanning ultraviolet-visible light diffuse reflection spectrophotometer (denoted as UV-vis DRS; Varian, USA) equipped with a labsphere diffuse reflectance accessory was performed to collect the UV-vis diffuse reflectance spectra. The adsorption spectra of N-TiO$_2$ and W + N codoped TNAs in the range of 300~800 nm were recorded with the UV-vis DRS facility at a scan rate of 600 nm min$^{-1}$.

*2.3. Photocatalytic Activity Measurements.* Photocatalytic activity of all samples was evaluated by monitoring the photocatalytic degradation of methylene blue (denoted as MB) in aqueous solution under visible light with vertical irradiation as light source. Briefly, to-be-tested N-TiO$_2$ and W-N-TiO$_2$-X (X = water, 0.01, 0.05, 0.5, 1, and 3) samples were dipped into 25 mL of MB solution with an initial concentration of 10 mg L$^{-1}$. The effective photocatalytic reaction area of all samples was 4 cm$^2$. A 300 W Xenon lamp (PLS-SEX300, Beijing Changtuo) was used as the light source. A cut-off filter was used to remove any radiation below 420 nm to ensure visible light irradiation only. The light intensity was measured by a photodetector, and the value is about 30 mW/cm$^2$ at 420 nm. Prior to irradiation, the MB solution containing to-be-tested sample was magnetically stirred in the dark for 30 min to establish adsorption-desorption equilibrium. During the photocatalytic reaction, the absorbance of MB solution at 664 nm was measured using an SP-2000 spectrophotometer at a time interval of 30 min. The decoloration rate of MB solution is calculated as $(C_0 - C)/C_0 \times 100\%$, where $C_0$ is the concentration of MB at adsorption–desorption equilibrium in the dark and $C$ is the concentration of MB upon completion of the photocatalytic reaction under visible light irradiation. In this way, the influence of adsorption amount of MB on its solution decoloration rate is excluded (different photocatalyst samples adsorb different amount of MB).

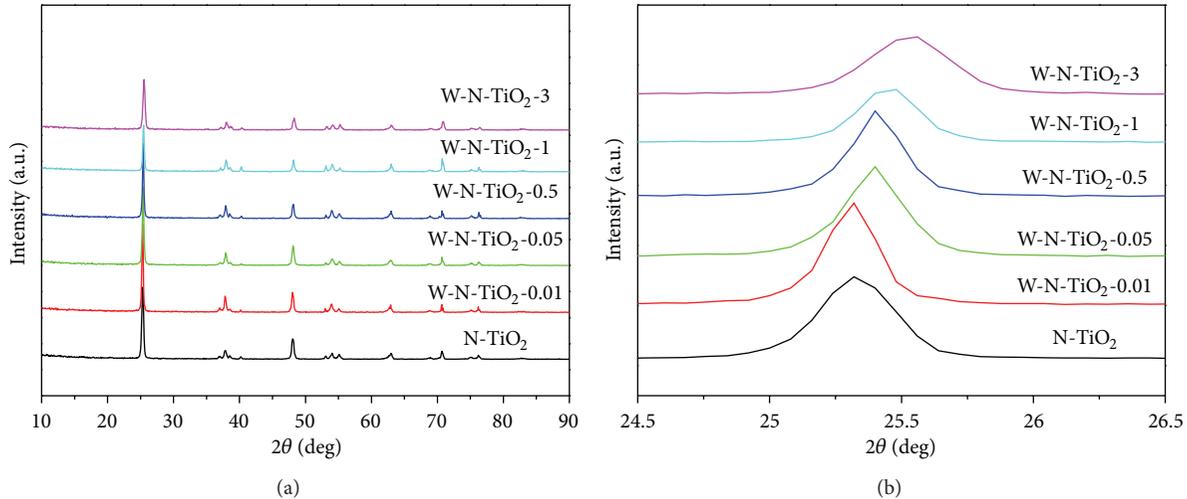

Figure 1: XRD patterns of N-TiO$_2$ and W-N-TiO$_2$-X (a), the enlarged XRD peaks of crystal plane (101) (b).

## 3. Results and Discussion

The phase structure, crystallite size, and crystallinity of TiO$_2$ play important roles in photocatalytic activity and photochemical properties. Many studies have confirmed that the anatase phase of TiO$_2$ exhibits higher photocatalytic activity than brookite or rutile phases [27]. Herein, XRD measurements were performed to investigate the changes of phase structure of TiO$_2$ nanotube arrays codoped with different ammonium tungstate addition. XRD patterns of various samples are shown in Figure 1. It can be seen that all the diffraction peaks of N-TiO$_2$ and W-N-TiO$_2$-X can be ascribed to anatase TiO$_2$, which indicates that W and N codoping has no effect on the crystal structure and phase composition of TNAs. It is worth noting that no WO$_3$ phase could be observed in all the XRD patterns, even in that of the sample with the W doping amount reaching 3%. Accordingly, we propose that tungsten ions may be incorporated into the titania lattice and replaced titanium ions to form W–O–Ti bonds.

From the enlarged image in Figure 1(b), it is interesting to note that W and N codoping causes the diffraction peak of anatase TiO$_2$ along (101) plane to shift towards higher values of 2θ. Moreover, slight enhancement of crystallinity properties can be observed when comparing the (101) peak intensities of the TNAs samples before and after the hydrothermal treatments. The XRD results imply that the Ti atoms in the anatase lattice of codoped TNAs samples may be substituted by W atoms, since W$^{6+}$ ions with a radius close to that of Ti$^{4+}$ are easy to enter into the lattices of TiO$_2$ and displace Ti$^{4+}$ ions, while nitrogen and tungsten have a synergistic doping effect [17, 21].

Figure 2 presents typical FESEM top view images of the N-TiO$_2$ and W-N-TiO$_2$-1 photocatalysts. Both images show the similar morphology of highly ordered nanotube arrays grown on the Ti substrate. SEM observation indicates that there is no apparent structural transformation of the TNAs samples after hydrothermal codoping process. The nanotubes of N-TiO$_2$ samples are open at the top end with an average diameter of 70 nm (Figure 2(a)). However, the diameter of nanotube in W-N-TiO$_2$-1 sample is found to be slightly increased to 80 nm. This difference of pore size is due to the enhanced crystallinity of the anatase nanotube walls after hydrothermal treatments in agreement with the XRD results. Yu et al. had reported that the nanotube array structures were completely destroyed after 180°C hydrothermal treatments with as-prepared TNAs samples due to the strongly enhanced anatase crystallinity and phase transformation from amorphous to anatase [28]. In our work, the as-prepared TNAs samples were calcinated at 500°C to realize the phase transformation from amorphous to anatase before subsequently hydrothermal process. The reported hydrothermally induced collapse was prevented with the simple calcination method. The calcined TNAs samples show better crystallinity with no apparent morphology change after hydrothermal codoping process.

XPS measurements are used to investigate the chemical states of W, N, Ti, and O in N-TiO$_2$ and W-N-TiO$_2$-1 samples. Figure 3(a) shows the high resolution XPS spectra for the N 1$s$ region of N-TiO$_2$ and W-N-TiO$_2$-1. The N 1$s$ binding energy peaks are broad, extending from 396 to 403 eV. The centre of the N 1$s$ peak locates at ca. 399.8 eV and 400.1 eV for N-$TiO_2$ and W-N-$TiO_2$-1 sample, respectively. These two peaks are much higher than that of typical binding energy of N 1$s$ (396.9 eV) in TiN [29], indicating that the N atoms in N-$TiO_2$ and W-N-$TiO_2$-1 interact strongly with O atoms [30]. The binding energy of 399.8 eV and 400.1 eV here is attributed to the oxidized nitrogen similar to $NO_x$ species, meaning Ti-N-O linkage possibly formed on the surface of N-$TiO_2$ and W-N-$TiO_2$-1 [13, 31, 32]. The N 1$s$ peak of N-$TiO_2$ at 399.8 eV is assigned to interstitial nitrogen species according to the previous report of Varghese et al. [26]. They suggested that the nitrogen incorporated within $TiO_2$ nanotubes during anodization is primarily supplied by $NH_4F$ or low extent of atmospheric nitrogen dissolved into the electrolyte. Furthermore, it is highly possible that some N-O species coordinate with W to form the W-N-O linkage on the surface of W-N-$TiO_2$-1. For the latter case, the electron density of N atoms in W-N-O is lower than that in Ti-N-O since the electronegativity of W is larger than that of Ti [13, 33]. It may explain why the binding energy of nitrogen surface species for W-N-$TiO_2$-1 (400.1 eV) is higher than that for N-$TiO_2$ (399.8 eV). High resolution W $4d$ spectra of N-$TiO_2$ and W-N-$TiO_2$-1 are presented in Figure 3(b). The W $4d$ peaks of W-N-$TiO_2$-1 at about 247.5 eV and 260.2 eV are assigned to $4d_{5/2}$ and $4d_{3/2}$ electronic states of $W^{6+}$, respectively [19, 20].

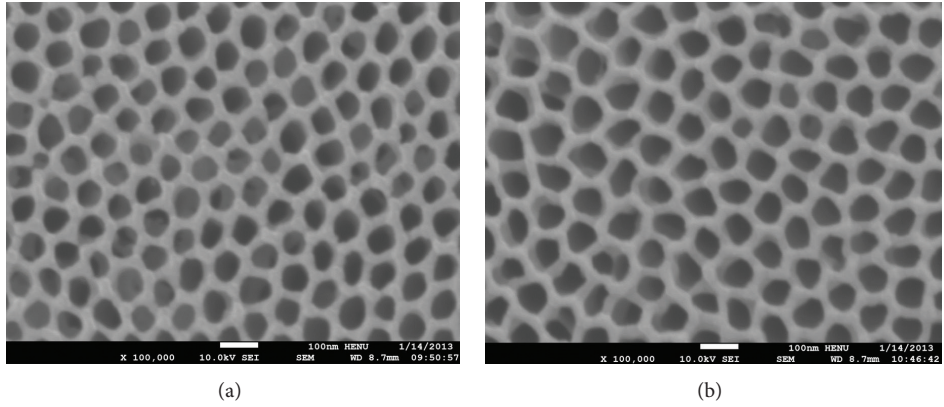

Figure 2: FESEM cross-sectional views of W and N codoped $TiO_2$ nanotube arrays: (a) N-$TiO_2$, (b) W-N-$TiO_2$-1.

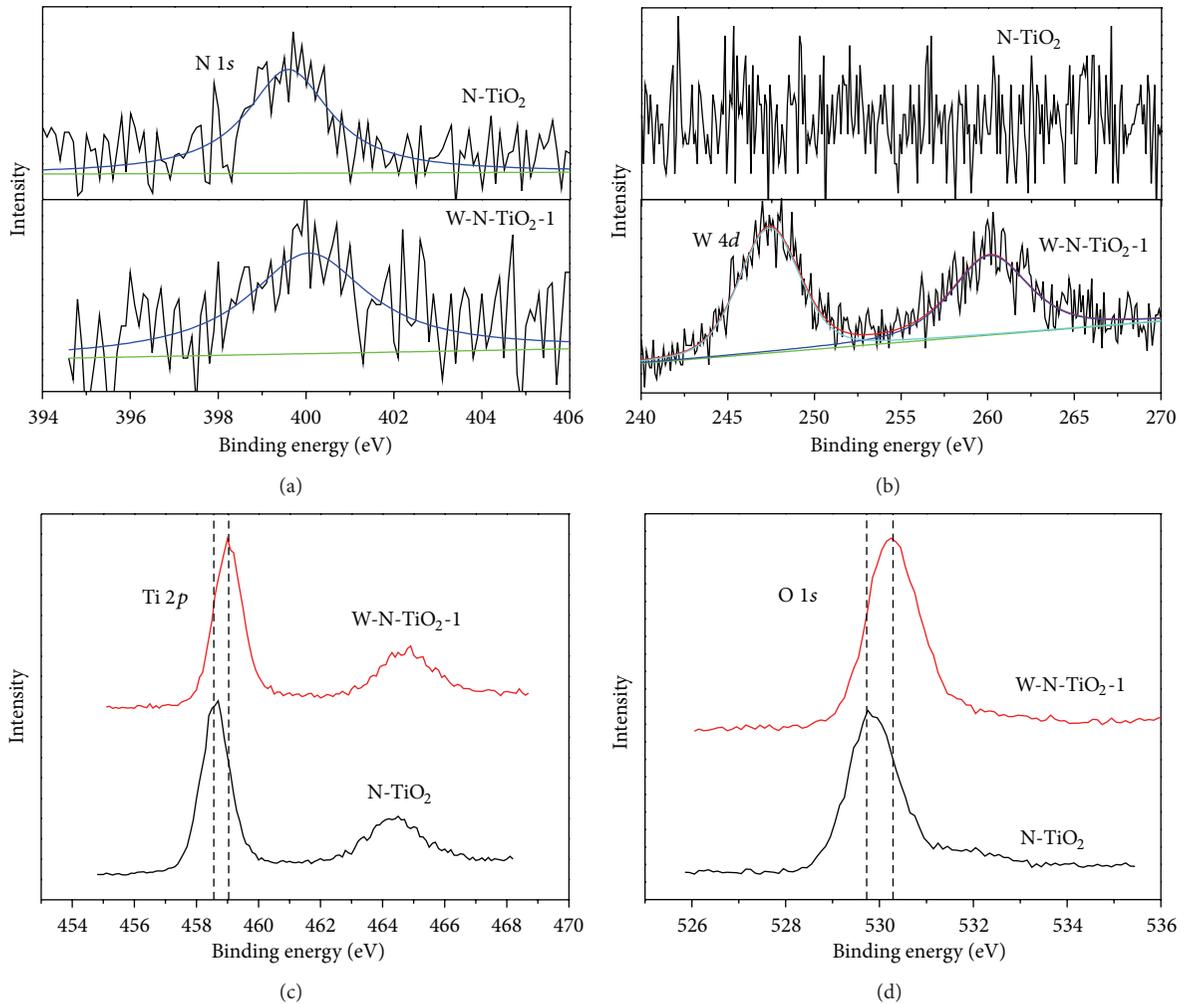

Figure 3: High resolution XPS spectra of (a) N 1$s$, (b) W 4$d$, (c) Ti 2$p$, and (d) O 1$s$ for N-$TiO_2$ and W-N-$TiO_2$-1 samples.

Figures 3(c) and 3(d) show high-resolution spectra of Ti $2p$, O $1s$ of N-$TiO_2$, and W-N-$TiO_2$-1. For N-$TiO_2$ sample, two peaks at 458.6 and 464.3 eV correspond to the Ti $2p_{3/2}$ and Ti2 $p_{1/2}$ states, indicating that Ti is 4+ valence. A strong peak centered at 529.8 eV is observed in the O $1s$ XPS spectra of N-$TiO_2$, attributing to bulk oxygen bonded to titanium. For the W and N codoped $TiO_2$ sample, both the Ti $2p$ and O $1s$ peaks are slightly shifted toward higher binding energy due to the doping of W in the $TiO_2$ lattice, which is in agreement with the result reported by Gong et al. [34]. All XPS results indicate that the $W^{6+}$ ions were doped into the bulk $TiO_2$ lattice by displacing $Ti^{4+}$ ions and forming the W–O–Ti bonding, which is in accordance with relevant XRD analytical results. According to the previous XPS and XRD results, it indicates that nitrogen is present as interstitial nitrogen in codoped TNAs samples, and W incorporates into $TiO_2$ lattice in substitutional mode.

The exact amounts of surface elements in all doped samples were studied by XPS analysis. The surface N/Ti and W/Ti ratio of all samples are calculated and listed in Table 1. It is found that the surfaces N/Ti and W/Ti ratio increased with the concentration of ammonium tungsten solution. Then, both the surfaces N/Ti and W/Ti ratio reaches their maximum value with the concentration of ammonium tungsten solution to 1% (theoretical doping concentration). However, the N/Ti and W/Ti ratio decreased clearly with further increasing the ammonium tungsten concentration to 3%. It indicates that the dopant concentration has an optimal value of about 1%. Above this optimal concentration, excess W and N ions cannot be doped in $TiO_2$ and show no contribution to the visible light absorption and subsequent photocatalytic activity. Moreover, the concentration of doped N increases with the codoping of W, which demonstrates that W and N codoping is in favor of enhancing the dopant solubility.

Figure 4 presents the UV-vis diffuse reflectance spectra of all samples. A relatively weak visible light absorption can be observed for the oxidized N-$TiO_2$ sample, which corresponds to a typical reported absorption feature of nitrogen-doped $TiO_2$ arising from the electron transition from surface state of $NO_x$ species to the conduction band of $TiO_2$ [13, 35]. Compared with that of N-$TiO_2$ sample, UV-vis spectra of all W and N codoped TNAs samples present red-shifted absorption edge and a strong absorption in the visible light region with typical extended absorption tail of about 600 nm. Moreover, for W and N codoped TNAs samples, the intensity of visible light absorption is enhanced with the increase of W and N concentration (the ammonium tungsten solution concentration from 0.01% to 1%). We consider that this absorption enhancement is due to the contribution of both interstitially doped N and substitutionally doped W. W and N incorporated in the lattice of TNAs show synergistic doping effect which favors to increase the solubility limits of W and N thereby greatly enhancing the visible light absorption along with the increase of the up-limits of W and N concentration [15, 36]. For the W-N-$TiO_2$-3 sample, the absorption edge is decreased because the superfluous W and N atoms cannot be doped into the lattice of $TiO_2$. Overall, the UV-vis DRS results indicate that nitrogen and tungsten codoped $TiO_2$ nanotube arrays are more sensitive to the visible light than N-$TiO_2$ samples.

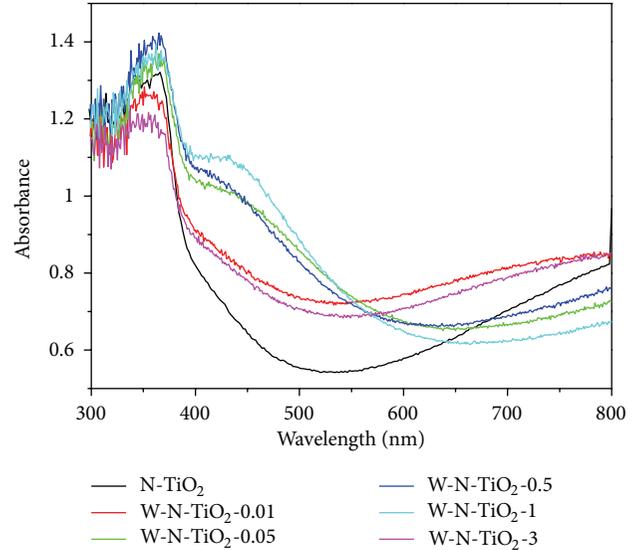

Figure 4: UV-vis diffuse reflectance spectra of N-$TiO_2$ and W-N-$TiO_2$-X samples.

The photocatalytic activities of N-$TiO_2$ and hydrothermal treated W-N-$TiO_2$-X samples were evaluated by monitoring the degradation of MB under visible light irradiation. To investigate the effect of hydrothermal treatment on the photocatalytic activity, the oxidized N-$TiO_2$ was firstly hydrothermally treated in pure water, and the obtained sample was denoted as W-N-$TiO_2$ water. An obviously enhanced photocatalytic activity was found for the W-N-$TiO_2$ water sample after hydrothermal treatment as shown in Figure 5. Many reports have shown that the photocatalytic activity of $TiO_2$ samples strongly depends on the preparing methods and posttreatments [36, 37]. Yu et al. had demonstrated an obvious increase of photocatalytic activity after thermal treatment of $TiO_2$ (P25) in pure water [36]. Recently, Dai et al. reported the enhanced photoelectrocatalytic activity of the $TiO_2$ nanotube arrays by hydrothermal modification [37]. They suggested that the enhanced photocatalytic activity was attributed to the formation of increasing number of hydroxyl groups on the surface of $TiO_2$ nanotube arrays after hydrothermal treatment. The increased hydroxyl groups would react with photoexcited holes on the surface of TiO$_2$ and produce hydroxyl radicals, which in turn are powerful oxidants in the photocatalytic reaction. In our work, before codoping, the number of hydroxyl groups on the surface of N-TiO$_2$ decreases due to the dehydration reaction during the higher temperature calcination process of oxidized TiO$_2$ nanotube arrays. When N-TiO$_2$ was codoped with W and N by hydrothermal treatment, Ti-OH leading to an enhanced photocatalytic activity can be regenerated on the surface of W-N-TiO$_2$-X samples due to the rupture of Ti–O–Ti bond.

Table 1: W/Ti and N/Ti ratio of W-N-TiO$_2$-X samples determined by XPS analysis.

| Sample | N-TiO$_2$ | W-N-TiO$_2$-0.01 | W-N-TiO$_2$-0.05 | W-N-TiO$_2$-0.5 | W-N-TiO$_2$-1 | W-N-TiO$_2$-3 |
|--------|-----------|------------------|------------------|-----------------|---------------|---------------|
| W/Ti   | 0         | 0.11             | 0.11             | 0.12            | 0.16          | 0.12          |
| N/Ti   | 0.02      | 0.13             | 0.16             | 0.19            | 0.22          | 0.14          |

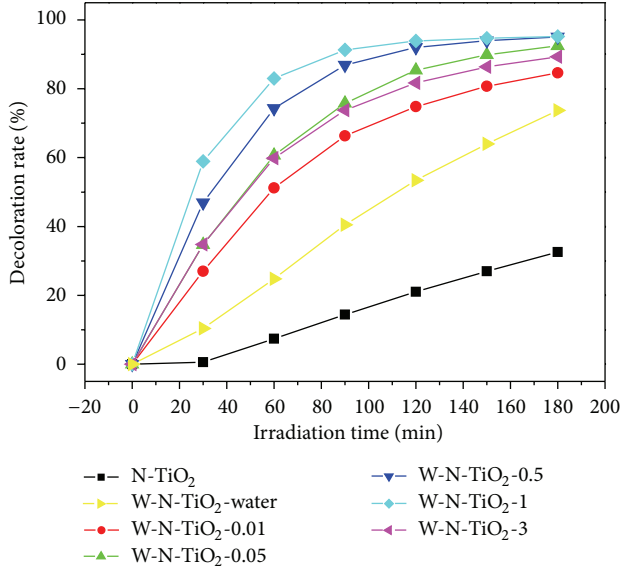

Figure 5: Photocatalytic degradation rate of MB in the presence of N-TiO$_2$ and W-N-TiO$_2$-X photocatalysts under visible light irradiation.

W and N codoping of N-TiO$_2$ samples was performed by the hydrothermal treatment in a series of concentrations of ammonium tungstate solution (concentration from 0.01% to 3%). As shown in Figure 5, W and N codoped TNAs possess much higher photocatalytic activity under visible light irradiation than N-TiO$_2$ sample and the W-N-TiO$_2$-water sample hydrothermal treated in pure water. Besides, the decoloration rate of MB for W and N codoped TNAs increases under visible light irradiation with the increase of dopant concentration. The decoloration rate reaches a maximum value for W-N-TiO$_2$-1, and it allows nearly complete elimination of MB within 3 h. Then, the photocatalytic activity for W-N-TiO$_2$-3 shows clear decrease due to its visible light absorption as discussed before.

The origin of the visible light photocatalytic activity of W and N codoped TNAs can be attributed to the interactions between N doping and W doping in the preparation processes. For the single N-doped sample (N-TiO$_2$) prepared by electrochemical anodization, isolated N 2$p$ states above the top of valence band are responsible for the red shifts in the optical absorption edge. After hydrothermal codoping with W and N, a hybridized state (composed of N 2$p$ orbitals and W 5$d$ states) is formed; in particular, the hybridized states are located mainly at the edge of the valence band, whereas other W 5$d$ states are at the edge of the conduction band. W 5$d$ states may contribute to the lowering of the energy levels of the N 2$p$ states, bringing the N states closer to the valence band and therefore enhancing mixing of N 2$p$ and O 2$p$ states in the valence band. Therefore, the addition of W to N-TiO$_2$ changes the character of N 2$p$ orbitals from isolated midgap states to N 2$p$ states mixed with O 2$p$ states [38]. At the same time, W 5$d$ states were located at the conduction band edge. Moreover, as electrons move from occupied W 5$d$ states to empty lower energy N 2$p$ states, a strong N–W bond forms in N and W codoped TNAs. This synergistic effect is beneficial for the increased doping amount of W and N in TiO$_2$ nanotube arrays. Consequently, the band gap energy of TiO$_2$ is significantly reduced, and its absorption in the visible light region is considerably increased due to modified conduction and valence band edges simultaneously. The band gap narrowing effect makes it feasible for W and N codoped TNAs to be activated by visible light with more electrons and holes being generated to participate in the photocatalytic reactions.

Moreover, W ions with changing valences in all codoped samples are considered to act as a temporary photogenerated electron or hole-trapping site thereby inhibiting the recombination of photogenerated charge carriers, prolonging their lifetime and improving the photocatalytic activity [17]. The detailed reaction steps are as follows:

$$\begin{aligned}
TiO_2 + h\nu &\longrightarrow e^- + h^+, \\
W^{6+} + e^- &\longrightarrow W^{5+}, \\
W^{5+} + O_2 &\longrightarrow W^{6+} + O_2^{-\bullet}, \\
W^{6+} + h^+ &\longrightarrow W^{7+}, \\
W^{7+} + OH^- &\longrightarrow W^{6+} + OH^\bullet.
\end{aligned} \quad (1)$$

When $W^{6+}$ ($d^5$) trap an electron, its electronic configuration is transferred to $d^6$, and if it traps a hole, its electronic configuration will be transferred to $d^4$ which is highly unstable. Therefore, to restore the stable electronic configuration, the trapped charge carrier tends to be transferred from $W^{5+}$ or $W^{7+}$ to the adsorbed $O_2$ or surface hydroxyl ($OH^-$) thereby regenerating $W^{6+}$. These newly produced active species (such as OH$^\bullet$ and O$_2^{-\bullet}$) are able to initiate the photocatalytic reactions.

## 4. Conclusions

W and N codoped TiO$_2$ nanotube arrays have been successfully synthesized by combining anodizing technique with hydrothermal method. W and N codoped TNAs samples show better crystallinity with no apparent morphology change after hydrothermal codoping process. XPS data reveal that, in as-prepared W and N codoped TNAs, nitrogen is present in the forms of Ti-N-O and W incorporates into TiO$_2$ lattice in substitutional mode. UV-vis DRS data show that W and N codoping was found to successfully extend the absorption of TiO$_2$ nanotube arrays into the whole visible light region. Increased number of hydroxyl groups produced by hydrothermal treatment shows an important role for the enhancement of photocatalytic activity. The synergetic doping effect between W and N plays a key role in producing new states, narrowing the band gap, and reducing the recombination effectively thereby greatly improving the visible light absorption and photocatalytic activity of TNAs. Besides, W ions with multiple valences in W-N-TiO$_2$-X samples can act as trapping sites to effectively decrease the recombination rate of electrons and holes, also resulting in improved photocatalytic activity of TNAs. Therefore, W and N codoped TiO$_2$ nanotube arrays with higher visible light absorption and photocatalytic activity are promising materials for visible light photocatalysts.

## Conflict of Interests

The authors declare that they do not have any commercial or associative interests that represents a conflict of interest in connection with the work submitted.

## Acknowledgments

The authors thank the National Natural Science Foundation of China (Grant no. 21203054) and the Science and Technology Department of Henan Province (Natural Science Project Grant no. 112300410171) for financial support.

## References

[1] A. Fujishima, X. T. Zhang, and D. A. Tryk, "TiO$_2$ photocatalysis and related surface phenomena," *Surface Science Reports*, vol. 63, no. 12, pp. 515–582, 2008.

[2] H. Tong, S. X. Ouyang, Y. P. Bi, N. Umezawa, M. Oshikiri, and J. H. Ye, "Nano-photocatalytic materials: possibilities and challenges," *Advanced Materials*, vol. 24, no. 2, pp. 229–251, 2012.

[3] X. B. Chen and S. S. Mao, "Titanium dioxide nanomaterials: synthesis, properties, modifications and applications," *Chemical Reviews*, vol. 107, no. 7, pp. 2891–2959, 2007.

[4] R. Asahi, T. Morikawa, T. Ohwaki, K. Aoki, and Y. Taga, "Visible-light photocatalysis in nitrogen-doped titanium oxides," *Science*, vol. 293, no. 5528, pp. 269–271, 2001.

[5] S. U. M. Khan, M. Al-Shahry, and W. B. Ingler Jr., "Efficient photochemical water splitting by a chemically modified n-TiO$_2$," *Science*, vol. 297, no. 5590, pp. 2243–2245, 2002.

[6] J. G. Yu, G. P. Dai, Q. J. Xiang, and M. Jaroniec, "Fabrication and enhanced visible-light photocatalytic activity of carbon self-doped TiO$_2$ sheets with exposed 001 facets," *Journal of Materials Chemistry*, vol. 21, no. 4, pp. 1049–1057, 2011.

[7] X. H. Tang and D. Y. Li, "Sulfur-doped highly ordered TiO$_2$ nanotubular arrays with visible light response," *Journal of Physical Chemistry C*, vol. 112, no. 14, pp. 5405–5409, 2008.

[8] M. H. Zhou and J. G. Yu, "Preparation and enhanced daylight-induced photocatalytic activity of C,N,S-tridoped titanium dioxide powders," *Journal of Hazardous Materials*, vol. 152, no. 3, pp. 1229–1236, 2008.

[9] Y.-C. Nah, I. Paramasivam, and P. Schmuki, "Doped TiO$_2$ and TiO$_2$ nanotubes: synthesis and applications," *ChemPhysChem*, vol. 11, no. 13, pp. 2698–2713, 2010.

[10] U. G. Akpan and B. H. Hameed, "The advancements in sol-gel method of doped-TiO$_2$ photocatalysts," *Applied Catalysis A*, vol. 375, no. 1, pp. 1–11, 2010.

[11] Y.-F. Li, D. H. Xu, J. I. Oh, W. Z. Shen, X. Li, and Y. Yu, "Mechanistic study of codoped titania with nonmetal and metal ions: a case of C + Mo Codoped TiO$_2$," *ACS Catalysis*, vol. 2, no. 3, pp. 391–398, 2012.

[12] J. L. Zhang, Y. M. Wu, M. Y. Xing, S. A. K. Leghari, and S. Sajjad, "Development of modified N doped TiO$_2$ photocatalyst with metals, nonmetals and metal oxides," *Energy and Environmental Science*, vol. 3, no. 6, pp. 715–726, 2010.

[13] E. J. Wang, T. He, L. S. Zhao, Y. M. Chen, and Y. A. Cao, "Improved visible light photocatalytic activity of titania doped with tin and nitrogen," *Journal of Materials Chemistry*, vol. 21, no. 1, pp. 144–150, 2011.

[14] X. X. Yang, C. D. Cao, K. Hohn et al., "Highly visible-light active C- and V-doped TiO$_2$ for degradation of acetaldehyde," *Journal of Catalysis*, vol. 252, no. 2, pp. 296–302, 2007.

[15] H. L. Liu, Z. H. Lu, L. Yue et al., "(Mo + N) codoped TiO$_2$ for enhanced visible-light photoactivity," *Applied Surface Science*, vol. 257, no. 22, pp. 9355–9361, 2011.

[16] Y. F. Shen, T. Y. Xiong, T. F. Li, and K. Yang, "Tungsten and nitrogen co-doped TiO$_2$ nano-powders with strong visible light response," *Applied Catalysis B*, vol. 83, no. 3-4, pp. 177–185, 2008.

[17] J. X. Li, J. H. Xu, W.-L. Dai, H. Li, and K. Fan, "One-pot synthesis of twist-like helix tungsten-nitrogen-codoped titania photocatalysts with highly improved visible light activity in the abatement of phenol," *Applied Catalysis B*, vol. 82, no. 3-4, pp. 233–243, 2008.

[18] A. Kubacka, B. Bachiller-Baeza, G. Colón, and M. Fernández-García, "W, N-codoped TiO$_2$-anatase: a sunlight-operated catalyst for efficient and selective aromatic hydrocarbons photo-oxidation," *Journal of Physical Chemistry C*, vol. 113, no. 20, pp. 8553–8555, 2009.

[19] A. Kubacka, B. Bachiller-Baeza, G. Colón, and M. Fernández-García, "Doping level effect on sunlight-driven W,N-co-doped TiO$_2$-anatase photo-catalysts for aromatic hydrocarbon partial oxidation," *Applied Catalysis B*, vol. 93, no. 3-4, pp. 274–281, 2010.

[20] A. Kubacka, G. Colón, and M. Fernández-García, "N- and/or W-(co)doped TiO$_2$-anatase catalysts: effect of the calcination treatment on photoactivity," *Applied Catalysis B*, vol. 95, no. 3-4, pp. 238–244, 2010.

[21] C. W. Lai and S. Sreekantan, "Study of $WO_3$ incorporated C-$TiO_2$ nanotubes for efficient visible light driven water splitting performance," *Journal of Alloys and Compounds*, vol. 547, pp. 43–50, 2013.

[22] P. Roy, S. Berger, and P. Schmuki, "$TiO_2$ nanotubes: synthesis and applications," *Angewandte Chemie*, vol. 50, no. 13, pp. 2904–2939, 2011.

[23] G. K. Mor, O. K. Varghese, M. Paulose, K. Shankar, and C. A. Grimes, "A review on highly ordered, vertically oriented $TiO_2$ nanotube arrays: fabrication, material properties, and solar energy applications," *Solar Energy Materials and Solar Cells*, vol. 90, no. 14, pp. 2011–2075, 2006.

[24] G. P. Dai, J. G. Yu, and G. Liu, "Synthesis and enhanced visible-light photoelectrocatalytic activity of p-N junction $BiOI/TiO_2$ nanotube arrays," *Journal of Physical Chemistry C*, vol. 115, no. 15, pp. 7339–7346, 2011.

[25] G. T. Yan, M. Zhang, J. Hou, and J. J. Yang, "Photoelectrochemical and photocatalytic properties of N + S co-doped $TiO_2$ nanotube array films under visible light irradiation," *Materials Chemistry and Physics*, vol. 129, no. 1-2, pp. 553–557, 2011.

[26] O. K. Varghese, M. Paulose, T. J. LaTempa, and C. A. Grimes, "High-rate solar photocatalytic conversion of $CO_2$ and water vapor to hydrocarbon fuels," *Nano Letters*, vol. 9, no. 2, pp. 731–737, 2009.

[27] J. G. Yu and B. Wang, "Effect of calcination temperature on morphology and photoelectrochemical properties of anodized titanium dioxide nanotube arrays," *Applied Catalysis B*, vol. 94, no. 3-4, pp. 295–302, 2010.

[28] J. G. Yu, G. P. Dai, and B. Cheng, "Effect of crystallization methods on morphology and photocatalytic activity of anodized $TiO_2$ nanotube array films," *Journal of Physical Chemistry C*, vol. 114, no. 45, pp. 19378–19385, 2010.

[29] N. C. Saha and H. G. Tompkins, "Titanium nitride oxidation chemistry: an x-ray photoelectron spectroscopy study," *Journal of Applied Physics*, vol. 72, no. 7, pp. 3072–3079, 1992.

[30] M. Sathish, B. Viswanathan, R. P. Viswanath, and C. S. Gopinath, "Synthesis, characterization, electronic structure, and photocatalytic activity of nitrogen-doped $TiO_2$ nanocatalyst," *Chemistry of Materials*, vol. 17, no. 25, pp. 6349–6353, 2005.

[31] Q. C. Xu, D. V. Wellia, R. Amal, D. W. Liao, S. C. J. Loo, and T. T. Y. Tan, "Superhydrophilicity-assisted preparation of transparent and visible light activated N-doped titania film," *Nanoscale*, vol. 2, no. 7, pp. 1122–1127, 2010.

[32] X. B. Chen, Y. B. Lou, A. C. S. Samia, C. Burda, and J. L. Gole, "Formation of oxynitride as the photocatalytic enhancing site in nitrogen-doped titania nanocatalysts: comparison to a commercial nanopowder," *Advanced Functional Materials*, vol. 15, no. 1, pp. 41–49, 2005.

[33] M. X. Sun and X. L. Cui, "Anodically grown Si-W codoped $TiO_2$ nanotubes and its enhanced visible light photoelectrochemical response," *Electrochemistry Communications*, vol. 20, pp. 133–136, 2012.

[34] J. Y. Gong, W. H. Pu, C. Z. Yang, and J. L. Zhang, "Novel one-step preparation of tungsten loaded $TiO_2$ nanotube arrays with enhanced photoelectrocatalytic activity for pollutant degradation and hydrogen production," *Catalysis Communications*, vol. 36, pp. 89–93, 2013.

[35] H. Q. Sun, Y. Bai, W. Q. Jin, and N. P. Xu, "Visible-light-driven $TiO_2$ catalysts doped with low-concentration nitrogen species," *Solar Energy Materials and Solar Cells*, vol. 92, no. 1, pp. 76–83, 2008.

[36] J. G. Yu, H. G. Yu, B. Cheng, M. Zhou, and X. Zhao, "Enhanced photocatalytic activity of $TiO_2$ powder (P25) by hydrothermal treatment," *Journal of Molecular Catalysis A*, vol. 253, no. 1-2, pp. 112–118, 2006.

[37] G. P. Dai, S. Q. Liu, T. X. Luo, S. Wang, and A. Z. Hu, "Hydrothermal treatment and enhanced photoelectrocatalytic activity of anatase $TiO_2$ nanotube arrays," *Chinese Journal of Inorganic Chemistry*, vol. 28, pp. 1617–1622, 2012.

[38] R. Long and N. J. English, "First-principles calculation of nitrogen-tungsten codoping effects on the band structure of anatase-titania," *Applied Physics Letters*, vol. 94, no. 13, Article ID 132102, 2009.